\documentclass[prb,twocolumn,groupedaddress,showpacs]{revtex4}
\usepackage{amsmath,amssymb,graphicx}

\begin{document}

\title{Influence of elastic deformation of porous materials in adsorption-desorption process. A thermodynamic approach.}

\author{Annie Grosman and Camille Ortega}

\affiliation{Institut des Nanosciences de Paris (INSP), Universit\'e Paris 6, UMR-CNRS 75-88, Campus Boucicaut, 140 rue de Lourmel, 75015 Paris, France}
\email{annie.grosman@insp.jussieu.fr}

\date{\today}
\begin{abstract}
It has been known for a long time that the adsorption and condensation of gas cause elastic deformation of the porous matrix. The reversible formation of an adsorbed film, which precedes capillary condensation, results in an extension of the porous material while, in the hysteresis region, the negative liquid pressures under the concave menisci contract the porous matrix. The elastic deformation exhibits an hysteresis loop in the same pressure region as the adsorption phenomenon. These deformations have been neglected in practically all the theoretical treatments of adsorption. There are two reasons for this. First, the deformations are small in magnitude and were supposed to have small effects on the adsorption process, and second, no experiment has contradicted the existing model according to which, in systems where the pores interact, the source of interactions is pore-pore intersections. They are the experimental results obtained in SBA-15 and p$^{+}$-type porous silicon, systems in which the pores interact whereas they are not connected, which lead us to question these models and consider the elastic deformation of the pore walls as a possible coupling parameter. Based on the experimental work of C. H. Amberg and R. McIntosh [Can. J. Chem. $\bf{30}$, 1012 (1952)] who measured both the linear deformation of a porous glass rod and the adsorbed amount during isothermal adsorption of water, we develop a thermodynamic approach which includes the elastic energy of the solid. This approach is generic to all porous materials. In the region of reversible adsorption preceding the capillary condensation where the variation of the surface free energy can be deduced from adsorption data, the linear extension of the solid is proportional to the variation of the surface free energy and to the elastic constant of the solid. This is the crucial point of the paper: thermodynamics of adsorption is directly connected to the elastic properties of the porous solid. In the hysteresis region, this linear relation can be used to deduce the variation of the surface free energy from the deformation measurements, a calculation which cannot be done from adsorption data. We find that the surface free energy related to the elastic deformation is an important component of the total free energy. It is shown that the condensation branch represents the more stable states and that an energy barrier exists to evaporation which depends essentially on the elastic deformation. The pores interact through the deformation of the walls. Based on this interaction mechanism and on the shape of the scanning curves which are common to materials with interconnected pores such as porous glass or noninterconnected pores such as p$^{+}$-type porous silicon and SBA-15, we propose a scenario for the filling and emptying of these porous materials.
\end{abstract}



\maketitle

\section{Introduction}
This paper originates in the experimental observations we made when studying the adsorption of gas in p$^{+}$-type porous silicon \cite{CoasnePRL, Grosman08}. This porous system, formed on highly boron doped Si(100) substrate, is composed of straight pores, perpendicular to the substrate, separated from each other by single crystal Si walls of apparent constant thickness ($\sim$5 nm). The section of the pores by a plane parallel to the substrate is polygonal. The pore size distribution (PSD) is large, for example $13\pm6$~nm for a porosity of $50\%$. Except the presence of different facets on the Si walls which gives some roughness, the amplitude of which is smaller than the apparent wall thickness and a fortiori than the pore size, the pore section does not vary along the pore axis. The pores are not interconnected.

\begin{figure}
\resizebox{\columnwidth}{!}{\includegraphics{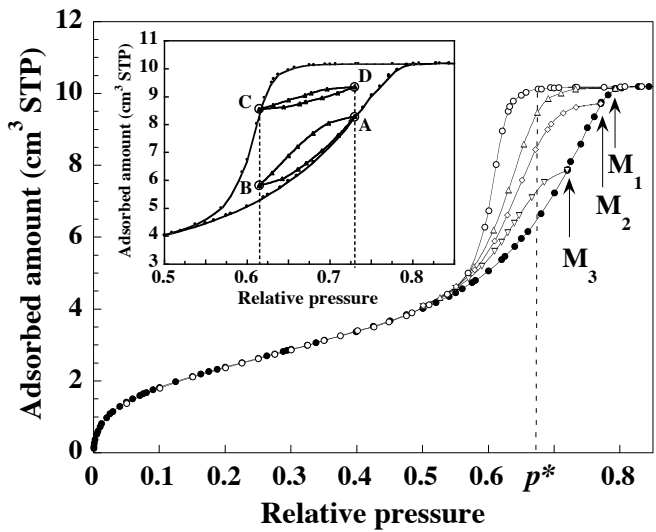}} 
\caption{Nitrogen adsorption isotherm at $77.4$~K for a p$^{+}$-type porous silicon layer of $50\%$ porosity exhibiting a large PSD ($13\pm6$~nm), together with three PDSCs starting from points $M_{1}$, $M_{2}$ and $M_{3}$ on the boundary condensation branch. $p^{*}$ is the critical pressure below which the fully filled system begins to empty. The inset shows a magnification of the hysteresis loop region and two subloops between the same pressure end points. The lack of congruence shows that the pores are non independent.}
\label{fig:hystbehavior}
\end{figure}

 At a first glance, this system can be hence classified as ordered porous system composed of independent pores. Nevertheless, we have shown that the hysteretic behavior of such a system, i.e. the shape of the main loop and of subloops inside the main loop, is characteristic of the presence of a strong interaction mechanism between the pores \cite{Grosman08}. In Fig.~\ref{fig:hystbehavior} we have represented the main features which characterize this system.
 
(i)- The boundary hysteresis loop, of type H2 in the International Union of Pur and Applied Chemists (IUPAC) classification \cite{Gregg82}, is asymmetrical with a broad condensation branch characteristic of the broad PSD and a steeper evaporation branch which suggests, but not proves, that, as soon as the pressure is reduced to a critical value $p^{*}$, the system begins to empty, leading to the emptying of the whole system in a avalanchelike manner.

(ii)- If the emptying of the system is commenced when it is not completely filled, at points $M_{1}$, $M_{2}$, $M_{3}$ on the boundary condensation branch for example, we obtain the so-called primary descending scanning curves (PDSCs). Along the PDSCs shown in Fig.~\ref{fig:hystbehavior}, evaporation occurs at pressure higher than $p^{*}$ which clearly shows that the evaporation pressure of a given pore depends on the state of the system.

(iii)- The two subloops between the same pressure end points shown in inset of Fig.~\ref{fig:hystbehavior} are not superimposable. For independent pores, according to Preisach model \cite{Preisach35}, they should be.

The pores of porous silicon interact strongly whereas they are not connected.

Similar observations have been made in other noninterconnected porous systems. MCM-41 and SBA-15 \cite{McNall01, Esparza04, Grosman05}, are composed of almost uniform cylindrical pores and exhibit hysteresis loops of type H1 with steep and parallel branches but, inside the main loop, the PDSCs are qualitatively similar to that represented in Fig.~\ref{fig:hystbehavior}. For SBA-15, we have shown that two subloops between the same pressure end points are not congruent which proves unambiguously that the pores of SBA-15 interact during the evaporation process \cite{Grosman05}. 

Surprising though it may seem, the hysteretic behavior represented in Fig.~\ref{fig:hystbehavior} is quite similar to that observed by Brown \cite{Brown63} in porous glass, a disordered porous material composed of cavities connected to one another by constrictions, a spongelike structure: the same hysteresis loop of type H2, the same PDSCs as porous silicon. Mason \cite{Mason} has developped a model, the so-called pore blocking/percolation model, to explain the family of PDSCs obtained by Brown. From this model, in porous materials where the pores interact, the source of interactions is thought to be pore-pore intersections \cite{Lilly1}.

The hysteretic behavior of SBA-15 has been compared to that of Kit-6, an ordered mesoporous silica which consists of a three-dimensional (3D) network of interconnected pores of almost cylindrical shape and same size. Morishige and coworkers \cite{Morishige07} noted that despite a large difference in porous structure, the shape and thermal behavior of the adsorption hysteresis, as well as the sorption scanning behavior for these two materials, are indistinguishable.

Finally, if we disregard the boundary hysteresis loops, the shape of which depends on the porous material, H1 for MCM-41, SBA-15 and Kit-6, H2 for porous glass and p$^{+}$-type porous silicon, the hysteretic features inside the main loop are qualitatively the same for all these porous systems and show that none of them is composed of independent pores. This suggests that the physical parameter which couple the pores is not interconnectivity.

The hysteresis loop and PDSCs shown in Fig.~\ref{fig:hystbehavior} have been qualitatively reproduced using mean-field density-functional theory or Monte Carlo calculations applied to a disordered lattice-gas model or to a simulated disordered matrix \cite{Ball89, Kierlik01, Woo01, Sarkisov02, Rosinberg03}. The calculations show the presence of multiple metastable states within the hysteresis region which are connected by PDSCs similar to that experimentally observed. As noted by the authors, these calculations reproduce the main features observed in disordered porous material without explicitly introducing pore blocking/percolation effects.

Following these calculations, it has been proposed that the hysteresis loop of type H2 observed in porous silicon is not due to "local pore blocking caused by individual constrictions of pore" but is rather the consequence of a strong disorder imposed by "large number of minor variations" in pore diameter\cite{Wallacher04} or by variation in the fluid/wall interaction along the pore \cite{Puibasset07}. We have already discussed this idea \cite{Grosman08} and concluded that the disorder in each pore of porous silicon cannot explain why the pores interact.

In the present paper, we will thus consider a new physical parameter, common to all these porous systems, which couples the pores during the adsorption-desorption process: the elastic deformation of the porous matrix.

The deformations of porous materials caused by adsorption and condensation of gas is an old subject \cite{Sereda67}. On the other hand, these deformations have been neglected in practically all the theoretical treatments of capillary condensation, that is, the condensed phase was always treated as a one component system of adsorbed molecules in the potential field of the adsorbent. However, as far back as 1956, Yates \cite{Yates56} noted that "the fact that the size changes do occur, even if small in magnitude, makes the assumption of an inert adsorbent, for physical adsorption, of very doubtful validity."

Hill first introduced, in the classical thermodynamics of adsorption, the porous matrix as a second element in addition to the adsorbate \cite{Hill}. Following this thermodynamic development, Quinn and McIntosh in a paper published in 1957, showed that the free energy of the porous glass-butane and porous glass-water \cite{Quinn57} systems depends importantly on the elastic deformation of the porous matrix.

Recently, Shen and Monson, based on the same thermodynamic property relationships as that developed by Hill, have made a Monte Carlo simulation study of gas adsorption in a semiflexible porous network \cite{Shen02}. They show that the flexible network makes a significant difference to both the adsorption and desorption isotherms. The evaporation, and to a lesser extent the condensation branches are shifted towards lower pressures. At the end of the paper, the authors noted that "the fact that the phase change involves a contraction of the solid volume may add to the barriers to the phase change and the ease with which hysteresis is exhibited by the system."

The thermodynamic approach developed in the present paper is quite different from that of Hill \cite{Hill}. The equilibrium condition for mass transfer is not given by the equality of the chemical potentials of the adsorbate and the vapor but depends on the variation of the free energy of the solid with the adsorbed amount. This equilibrium condition allows us to introduce, in the thermodynamic relationships, the elastic energy of the solid besides the surface free energy that is to connect the deformation of the solid to the variation of the surface free energy.

The paper is organized as follows. In Sec.~\ref{subsec:elasdef}, we present the results found in the literature dealing with the elastic deformation of porous materials. In Sec.~\ref{subsec:thermo}, we establish a thermodynamic approach of adsorption-desorption process, which takes into account the elastic deformation of the porous matrix. In this thermodynamic approach, we use the experimental work of Amberg and McIntosh. \cite{Amberg52} We estimate the numerical values of the different components of the free energy of the system adsorbate solid. In Sec.~\ref{subsec:qualitative}, we show how the pores can interact through the elastic deformation of the pore walls and we present a qualitative description of the hysteretic behavior shown in Fig.~\ref{fig:hystbehavior} which is common to all porous materials whether the pores are interconnected or not.
\section{Results and discussion}
\label{sec:result}
\subsection{Elastic deformation of porous materials}
\label{subsec:elasdef}
Adsorption and desorption of gas cause deformations of the porous matrix. These deformations have been studied in a number of systems such as porous glass \cite{Amberg52}, charcoal \cite{Wiig49}, silica aerogel \cite{Herman06} and porous silicon. \cite{Dolino96} The magnitude of the linear deformation depends on the stiffness of the bulk material, on the porosity, and on the properties of the adsorbed molecules. For adsorption of pentane, at ambient temperature, in p$^{+}$-type porous silicon, the variation of the lattice parameter ($\Delta a/a$), along the [100] axis perpendicular to the substrate, measured with regard to the Si substrate by x-ray diffraction observations, is a few $10^{-4}$. For the other cited porous materials, the order of magnitude of the linear deformation ($\Delta l/l$) is typically $10^{-3}$ for water in charcoal and porous glass and a few $10^{-2}$ for neon (T=$43$~K) in silica aerogel.

\begin{figure}
\resizebox{\columnwidth}{!}{\includegraphics{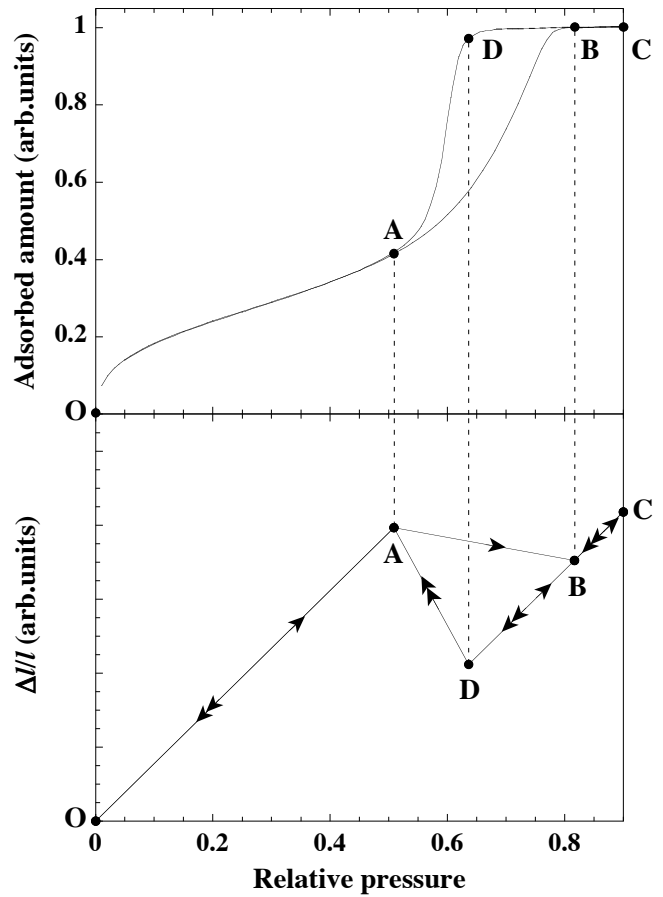}} 
\caption{Schematic representation of an adsorption isotherm presenting a hysteresis loop of type H2 and the corresponding typical linear extension ($\Delta l/l$) observed during the adsorption-desorption cycle.}
\label{fig:schemDeltaL}
\end{figure}

Despite the porous materials cited above having very different morphologies, the deformation during the condensation and evaporation of fluids presents common features. Figure~\ref{fig:schemDeltaL} represents schematically a hysteresis loop of type H2 and typical deformations observed in these porous materials during an adsorption-desorption cycle.

Along the reversible path OA preceding capillary condensation, the formation of an adsorbed film results either in a marked extension of the porous material (porous glass) or in small or insignificant changes in the other materials. The two-dimensional (2D) pressure exerted on the pore walls by the film is usually called spreading pressure. Between points A and B, where the condensation process takes place, a contraction is generally observed. During this step, the pores sequentially fill, from the smallest to the largest; menisci are presumably formed at the top of the filled pores, and the thickness of the adsorbed film increases reversibly in the empty pores leading to a further extension of the porous material. Therefore, the net contraction generally observed during capillary condensation is attributed to the negative pressure created under the concave menisci. At point B, all the pores are filled. Along BC, the upper region of the isotherm, a marked extension occurs, attributed to the vanishing of the negative pressure within the fluid as a result of the flattening of the concave menisci.

Conversely, along CD, marked contraction occurs due to large negative fluid pressure generated by the formation of concave menisci. In the region BCD, both adsorption and deformation data are generally reversible except for pentane in porous silicon \cite{Dolino96} for which the authors note that at the pentane saturation pressure ($60$~kPa) the extension has not reached the maximum value found for full immersion in the liquid. It is necessary to increase the pressure above the vapor saturation pressure by slightly heating the liquid reservoir above the temperature of the sample to obtain a full wetting. We have never observed such a phenomenon in the case of adsorption of N$_{2}$, Ar, or Kr in porous silicon: as soon as the pores are filled (point B), the adsorption-desorption path BCD is reversible. \cite{CoasnePRL, Grosman08}

Along DA, the pores empty by evaporation and the deformation recovers the initial value at point A. From A to O, the path is the reverse of that along OA.

Recently, it has been shown that, along CD, the lattice constant of solid Kr ($90$~K) confined to mesoporous spherical cavities increases as the vapor pressure is decreased. \cite{Morishige06} As the increase in the interatomic distance of the fluid is equivalent to an increase in the negative liquid pressure value, this supports the above explanation--according to which the contraction of the porous matrix along CD is due to the increased value of the negative liquid pressure under menisci.

The above results clearly display the deformation of the porous matrix during the reversible adsorption (OA), the capillary condensation (AB), and the evaporation (DA) processes. 

Our aim is now to establish the thermodynamic relationships which take into account the elastic deformation of the porous matrix during the adsorption-desorption process.
\subsection{Thermodynamics of adsorption in elastically deformed porous materials}
 \label{subsec:thermo}
\subsubsection*{\textit{\textbf{I-Reversible adsorption}}}
 \label{subsubsec:rev ads}
When a porous matrix of surface area $A$ is brought into contact with $N$ gas molecules contained in a volume $V$ at a constant temperature $T$, the thermodynamic potential of the adsorption system, including the porous solid (so), the adsorbate ($\sigma$) and the gas reservoir ($v$) is the free energy $F$:
\begin{equation}
F=F_{\sigma} +F_{\rm so}^{\rm bulk}+ F_{v}
\end{equation}
where $F_{\rm so}^{\rm bulk}$ and $F_{v}$ are the bulk free energies of the solid and of the vapor. $F_{\sigma}$, which includes the bulk of the adsorbate and the adsorbate-vapor and adsorbate-solid interfaces, depends, for isothermal conditions, on three extensive independent variables, $V_{\sigma}$, the volume of the adsorbed phase, $N_{\sigma}$, the number of adsorbed atoms, and $A$, the surface of the porous solid.
\begin{equation}
dF_{\sigma}=-P_{\sigma}dV_{\sigma}+\mu_{\sigma}dN_{\sigma}+\gamma dA
\end{equation}
where $\mu_{\sigma}$ is the chemical potential of the adsorbed atoms and $\gamma$ is the surface free energy per unit surface area.

The equilibrium condition for mass transfer, at constant $T$, $V_{v}$, $V_{\sigma}$, $N=N_{\sigma}+N_{v}$, and $A$ is
\begin{eqnarray}
\bigg(\frac{\partial F}{\partial N_{\sigma}}\bigg)\!\!\!&=&\!\!0\nonumber\\
\!\!\!&=&\!\!\bigg(\frac{\partial F_{\sigma}}{\partial N_{\sigma}}\bigg)_{V_{\sigma}, A}\!\!\!\!+\!\bigg(\frac{\partial F_{\rm so}^{\rm bulk}}{\partial N_{\sigma}}\bigg)_{N_{\rm so}}\!\!\!\!+\!\bigg(\frac{\partial F_{v}}{\partial N_{\sigma}}\bigg)_{V_{v}}\!\!\!
\end{eqnarray}
where $N_{so}$ is the total number of solid atoms.

As $\big(\partial F_{v}/\partial N_{\sigma}\big)_{V_{\sigma}, A}$= -$\mu_{v}$ since $dN=dN_{v}+dN_{\sigma}=0$, equilibrium condition (3) becomes
\begin{equation}
\mu_{\sigma}=\mu_{v} -\bigg(\frac{\partial F_{\rm so}^{\rm bulk}}{\partial N_{\sigma}}\bigg)_{N_{\rm so}}.
\end{equation}

If the properties of the bulk of the porous solid are supposed to be unaffected by the adsorbed molecules,
\begin{equation}
\bigg(\frac{\partial F_{\rm so}^{\rm bulk}}{\partial N_{\sigma}}\bigg)_{N_{\rm so}}=0
\end{equation}
and equilibrium condition (4) becomes
\begin{equation}
\mu_{\sigma}=\mu_{v}.
\end{equation}
The study of the adsorbed phase can be then disconnected from that of the porous solid as is generally done.

Now, if the elastic deformation of the solid is taken into account, $\big(\partial F_{\rm so}^{\rm bulk}/\partial N_{\sigma}\big)_{N_{\rm so}}\neq 0$ and the equilibrium condition is given by Eq.~(4), contrary to what has been assumed in all the papers where the deformation of the solid has been taken into account. \cite{Hill, Quinn57, Shen02} Actually, as is shown below, the shift ($\mu_{\sigma} -\mu_{v}$) is small, but even small it cannot be ignored if we want to treat consistently the influence of the elastic deformation on adsorption. Then, for isothermal conditions, $F_{\sigma}$ can be written as
\begin{equation}
dF_{\sigma}=-P_{\sigma}dV_{\sigma}+\mu_{v}dN_{\sigma}-\!\bigg(\frac{\partial F_{\rm so}^{\rm bulk}}{\partial N_{\sigma}}\bigg)_{N_{\rm so}}\!\!dN_{\sigma}+{\gamma} dA.
\end{equation}
For a given porous solid, $A$~=~$aN_{\rm so}$ where, let us point out it, $N_{\rm so}$ is the total number of solid atoms and not the number of surface atoms. Equation (7) becomes
\begin{equation}
dF_{\sigma}=-P_{\sigma} dV_{\sigma}+\mu_{v}dN_{\sigma}-\!\bigg(\frac{\partial F_{\rm so}^{\rm bulk}}{\partial N_{\sigma}}\bigg)_{N_{\rm so}}\!\!\!\!dN_{\sigma}+{\gamma}^{'}\!dN_{\rm so}
\end{equation}
where ${\gamma}^{'}=a{\gamma}$ is the surface free energy per solid atom.

It is convenient to introduce in Eq.~(8), instead of the total free energy of the solid, the change of its free energy with adsorption, which is nothing other than the elastic energy stored in the solid during the adsorption process, $F_{\rm so}^{el}$: 
\begin{equation}
F_{\rm so}^{el} =F_{\rm so}^{\rm bulk}-F^{\rm * bulk}_{\rm so},
\end{equation}
where $F^{\rm *bulk}_{\rm so}(T,~V^{*}_{\rm so},~N_{\rm so})$ is the bulk free energy of the porous matrix without adsorbate (the empty matrix) but at the same $T$ and $P$ as $F_{\rm so}^{\rm bulk}(T,~V_{\rm so},~N_{\rm so})$. $V^{*}_{\rm so}$ and $V_{\rm so}$ are the volumes of the porous solid before and after adsorption. $F_{\rm so}^{el}$ can be expressed as a function of the independent variables ($T$,~$N_{\sigma}$,~$N_{\rm so}$) instead of ($T$,~$V_{\rm so}$,~$V_{\rm so}^{*}$,~$N_{\rm so}$). Note that $N_{\sigma}$ is not an extensive variable for $F_{\rm so}^{el}$ in contrast to $N_{\rm so}$:
\begin{equation}
F_{\rm so}^{el} =N_{\rm so}~f_{\rm so}^{el}\,(T, N_{\rm{\sigma}}),
\end{equation}
where $f_{\rm so}^{el}\,(T, N_{\rm{\sigma}})$ is the elastic energy per solid atom. For isothermal conditions, we get
\begin{equation}
dF_{\rm so}^{el} =\bigg(\frac{\partial F_{\rm so}^{el}}{\partial N_{\sigma}}\bigg)_{N_{\rm so}}dN_{\sigma}+f_{\rm so}^{el}\,dN_{\rm so}.
\end{equation}
As $(\partial F_{\rm so}^{el}/\partial N_{\sigma})_{N_{\rm so}}$=$(\partial{F}_{\rm so}^{\rm bulk}/\partial N_{\sigma})_{N_{\rm so}}$, equilibrium condition (4) becomes
\begin{equation}
\mu_{\sigma}=\mu_{v} -\bigg(\frac{\partial F_{\rm so}^{el}}{\partial N_{\sigma}}\bigg)_{N_{\rm so}}
\end{equation}
and Eq.~(11) can be rewritten as
\begin{equation}
dF_{\rm so}^{el}=(\mu_{v}-\mu_{\sigma})dN_{\sigma}+f_{\rm so}^{el}\,dN_{\rm so}.
\end{equation}
Finally, from Eqs.~(8) and (11), we obtain
\begin{equation}
d(F_{\sigma} + F_{\rm so}^{el})=-P_{\sigma} dV_{\sigma}+\mu_{v}dN_{\sigma}+\Psi dN_{\rm so},
\end{equation}
where we have put
\begin{equation}
\Psi=f_{\rm so}^{el}+\gamma^{'}.
\end{equation}

$\Psi$ can be considered as an "effective" surface free energy per solid atom. After integrating and by introducing the new notation $F_{\sigma}+ F_{\rm so}^{el}=F_{\sigma,\rm so}$, Eq.~(14) becomes
\begin{equation}
F_{\sigma,\rm so}= - P_{\sigma}V_{\sigma}+\mu_{v}N_{\sigma}+\Psi N_{\rm so}.
\end{equation}

\textit{a. Relationship between $\Psi$ and the elastic deformation.} The Gibbs-Duhem relationship is, for isothermal conditions,
\begin{equation}
-V_{\sigma}dP_{\sigma}+N_{\sigma}d\mu_{v}+N_{\rm so}d\Psi=0.
\end{equation}
Assumming that the vapor follows the perfect gas law, and that $P_{\sigma}\approx P$, Eq.~(17) becomes
\begin{equation}
N_{\rm so}\int d\Psi= \int N_{\sigma}(v_{\sigma}-k_{B} T/P)dP,
\end{equation}
where $v_{\sigma}$, the volume occupied by an adsorbed molecule, is much smaller than $k_{B}T/P$, the volume occupied by a vapor molecule, and can be neglected. $P_{\sigma}$ is in general different from the vapor pressure $P$ since in porous materials the adsorbate-vapor interface is curved but, as long as we are concerned by the reversible adsorption of a few monolayers before the capillary condensation takes place, the term $P_{\sigma}V_{\sigma}$ is much smaller than $\gamma A$ and can be neglected: \cite{Guggenheim39}
\begin{equation}
N_{\rm so}\int d\Psi=-\int N_{\sigma}\,k_{B} T\,\frac{dP}{P}.
\end{equation}
Equation (19) gives $\Psi$ as a function of adsorption data.

At this stage, it is noteworthy that the calculation of $\Psi$ from adsorption data does not provide any information about the magnitude of the porous solid deformation. Independent measurements are hence necessary.

Some investigators \cite{Bangham30, Amberg52} have measured the deformation of a porous glass rod and have observed a linear variation of the relative extension of the rod length $\big(\Delta l/{l}\big)$ with $\int N_{\sigma}k_{B}T\,dP\!/\!P$\,:
\begin{equation}
\int _{0}^{P}N_{\sigma}\,k_{B}T\,\frac{dP}{P} = k \bigg(\frac{l-l_{0}}{l_{0}}\bigg)
\end{equation}
where $l_{0}$ and $l$ are the rod length before and after adsorption and $k$, a proportionality factor. 
\begin{figure}
\resizebox{\columnwidth}{!}{\includegraphics{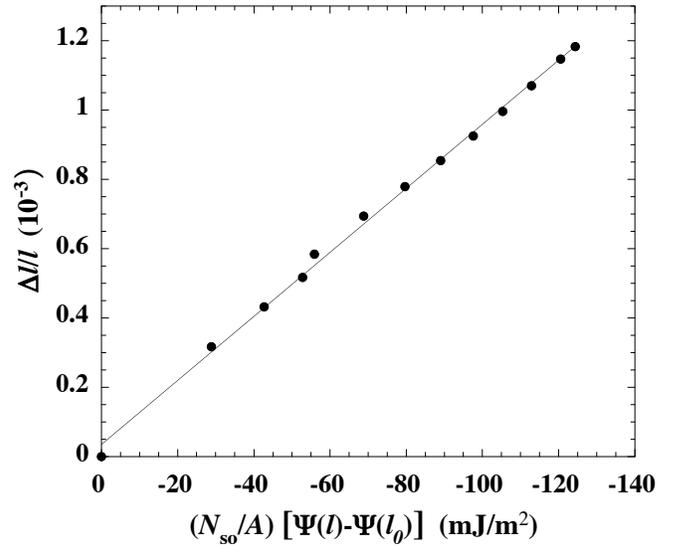}} 
\caption{Linear extension ($\Delta l/l$) of a porous glass rod measured by Amberg and McIntosh (Ref.~\cite{Amberg52}) during water adsorption at $18.75^\circ$C as a function of $(N_{\rm so}/A)\big[\Psi(l)-\Psi(l_{0})\big]$, where $N_{\rm so}$ is the total number of solid atoms, $A$, the surface area and $\big[\Psi(l)-\Psi(l_{0})\big]$, the variation of the "effective" surface free energy per solid atom as defined in Eq.~(15) and calculated according to Eq.~(19). $l_{0}$ is the length of the porous rod before adsorption. The surface area equals $129$~m$^2$/g. The regression coefficient of the linear fit equals 0.999.}
\label{fig:AmbergGamma}
\end{figure}
\begin{figure}
\resizebox{\columnwidth}{!}{\includegraphics{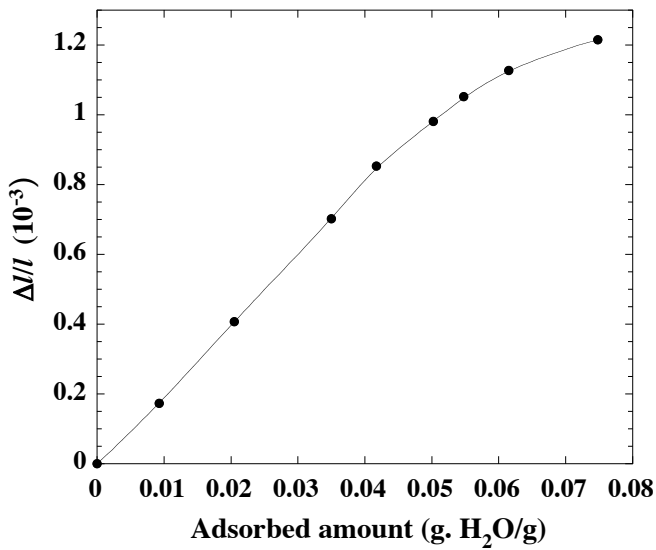}} 
\caption{Linear extension ($\Delta l/l$) of a porous glass rod measured by Amberg and McIntosh (Ref.~\cite{Amberg52}) during water adsorption at $18.75^\circ$C as a function of the adsorbed amount.}
\label{fig:AmbergAds}
\end{figure}
From Eq.~(19) and experimental relation (20), we get 
\begin{equation}
N_{\rm so}\big[\Psi(l)-\Psi(l_{0})\big]=- k \bigg(\frac{l-l_{0}}{l_{0}}\bigg).
\end{equation}

Figure~\ref{fig:AmbergGamma} shows for example the data obtained by Amberg and McIntosh \cite{Amberg52} for adsorption of water in porous glass. Similar results have been obtained in the case of butanol adsorption on exfoliated graphite. \cite{Dash76} For further discussions, we have also represented in Fig.~\ref{fig:AmbergAds} the relative extension ($\Delta l/l$) as a function of the adsorbed amount.

\textit{b. Relationship between $k$ and the elastic constants of the porous matrix.} For small deformation, $F_{\rm so}^{el}$ can be represented by the equation
\begin{equation}
F_{\rm so}^{el}= \frac{1}{2}\,C\,V_{\rm so}^{*}\,\bigg(\frac{l-l_{0}}{l_{0}}\bigg)^{2},
\end{equation}
where $C$ is a constant which depends on the elastic constants of the porous solid, i.e., Young's modulus and Poisson's ratio and also on the geometry of the material. In adsorption experiment, if the porous material is isotropic, the stresses and the deformations are isotropic. In this condition, the classical theory of elasticity gives $C=9~K$, where $K$ is the bulk modulus. 

Equation (16) can be written in the form:
\begin{eqnarray}
F_{\sigma,\rm so}&=&-P_{\sigma} V_{\sigma}+{\mu_{v} N_{\sigma}}\nonumber\\
&+&\frac{1}{2}\,C\,N_{\rm so}\,v_{\rm so}^{*}\,\bigg(\frac{l-l_{0}}{l_{0}}\bigg)^{2}+\gamma A.
\end{eqnarray}
The minimisation of $F_{\sigma,\rm so}$ with respect to $l$, keeping constant all the other variables, i.e., $V_{\sigma}$, $N_{\sigma}$ and also $l_{0}$ (that is $N_{\rm so}$), yields
\begin{equation}
\frac{\partial F_{\sigma, \rm so}}{\partial l}=0= C\,V_{\rm so}^{*}\,\frac{(l-l_{0})}{l_{0}^{2}}+\frac{\partial(\gamma A)}{\partial l}.
\end{equation}

$l(\partial/\partial l)= \alpha A(\partial/\partial A)$ where $\alpha$ is a coefficient, the value of which depends whether or not the porous material is isotropic. If the material is isotropic, as it is always the case for powders, $dA/A=2dl/l$ and $\alpha=2$. In the case of porous silicon, the porous layer is supported by the substrate and the planes perpendicular to the interface are constrained to have the same interatomic spacing as that of the substrate so that the transverse deformations can be neglected and $\alpha=1$. We get:
\begin{equation}
\alpha A\frac{\partial(\gamma A)}{\partial A}
=l_{0}\,\frac{\partial(\gamma A)}{\partial l}.
\end{equation}
The partial derivative, $\partial (\gamma A)/\partial A$\,, is directly related to the spreading pressure, $\Pi$, exerted by the adsorbed film on the substrate. Thus, equilibrium condition (24) leads to
\begin{equation}
\Pi=\frac{\partial (\gamma A)}{\partial A}=- C\,\frac{V_{\rm so}^{*}}{\alpha A}\,\frac{(l-l_{0})}{l_{0}},
\end{equation}
which is Hooke's law. The relation between the spreading pressure $\Pi$ and the surface free energy per unit area $\gamma$ follows from Eq.~(25):
\begin{equation}
\Pi=\frac{\partial(\gamma A )}{\partial A}
=\gamma + A\frac{\partial \gamma}{\partial A} =\gamma +\frac{l_{0}}{\alpha}\frac{\partial \gamma}{\partial l} 
\end{equation}
or
\begin{equation}
\Pi=\gamma +\frac{\partial \gamma}{\partial (\alpha \frac{l}{l_{0}})},
\end{equation}
which is a relation analogous to that established by Shuttleworth. \cite{Shuttleworth50} Similar calculations have been done by Dash \textit{et al.} \cite{Dash78} and by Thibault \textit{et al.} \cite{Thibault95} to determine the relation between the deformation of a porous solid and the variation of the surface free energy. 

The fact that $N_{\rm so}\Psi$ varies linearly with $\big(\Delta l/{l}\big)$ as shown by Eq.~(21), indicates that $f_{\rm so}^{el}\ll\gamma^{'}$ [see Eq.~(15)]. In this condition, as $N_{\rm so}\gamma^{'}=\gamma A$, Eq.~(21) becomes
\begin{equation}
\gamma(l)-\gamma(l_{0})\approx-\frac{k}{A} \frac{(l-l_{0})}{l_{0}}.
\end{equation}
Hence,
\begin{equation}
\frac{\partial \gamma}{\partial (\frac{l}{l_{0}})}= -\frac{k}{ A}
\end{equation}
and from the above equations, we get
\begin{equation}
k= \frac{C\,V_{\rm so}^{*}}{\alpha}
\end{equation}
and
\begin{equation}
\Pi=\gamma (l)-\gamma (l_{0})= -\frac{CV_{\rm so}^{*}}{\alpha A}\,\frac{(l-l_{0})}{l_{0}}.
\end{equation}
The spreading pressure $\Pi$ exerted by the adsorbed molecules on the porous matrix is hence equal to the variation of the surface free energy. The difference between these two quantities is discussed in a note. \cite{Dash78}

Equation (31), applied to an isotropic material, gives $k=9KV_{\rm so}^{*}/2$, a relation identical to that found by Scherer \cite{Scherer86} for adsorption on a spherical material.

Equation (32) is the crucial point of our problem. The variation of the surface free energy is directly connected to the variation of the elastic energy of the porous solid. This is an old result which has never been taken into account in the physics of adsorption.

\textit{c. Numerical estimations.} The comparison of elastic constants deduced from adsorption measurement (internal stress) and from external loading is a complicated task since it requires geometrical model describing the porous solids and is not the subject of this paper. We only give here an estimation of Young's modulus, $E$, deduced from the data represented in Fig.~\ref{fig:AmbergGamma} where $k=1.39\times10^{4}$~J/g.adsorbent. Assuming that porous glass is isotropic, we have $k=9KV_{\rm so}^{*}/2$ and $K=E/3(1-2\nu)$, where Poisson's coeficient $\nu$ equals $0.2$ to $0.25$. \cite{Scherer86} Taking a density of $2.7$~g/cm$^{3}$, we find $\simeq14$~GPa. If, on the other hand, the tranverse deformations are neglected as has been done by Amberg \textit{et al.}, \cite{Amberg52} $k=EV_{\rm so}^{*}$ which gives $E=37.7$~GPa. These values are of the same order of magnitude as those found by Scherer \cite{Scherer86} using sonic resonance namely in the range ($15-30$~GPa).

Now, compare the elastic energy stored in the porous matrix to the corresponding variation of the surface free energy. Equations (22) and (32) show that the ratio $F_{\rm so}^{el}/\gamma A$ is of the order of ($\Delta l/l$). More precisely, in Fig.~\ref{fig:AmbergGamma}, the last measurement point, for example, corresponds to a variation of surface free energy equal to $125$~mJ/m$^{2}\times129$~m$^{2}$/g=$16$~J/g. The corresponding extension is $(\Delta l/l)=1.19\times10^{-3}$ so that $F_{\rm so}^{el}=70\times10^{-3}$~J/g.

$F_{\rm so}^{el}$ is hence a small perturbation of $F_{\sigma,\rm so}$ but even small, it cannot be ignored if we want to understand how the surface free energy is connected to the relative deformation ($\Delta l/l$).

Finally, we can estimate the difference $(\mu_{v}-\mu_{\sigma})$. From Eq.~(13) we get, at constant $N_{\rm so}$ as it is the case in adsorption experiments, 
\begin{equation}
\int(\mu_{v}-\mu_{\sigma})\,dN_{\sigma}=F_{\rm so}^{el}\ll \gamma A,
\end{equation}
which means that $(\mu_{v}-\mu_{\sigma})$ is small compared to each of the two terms.

\textit{d. Summary.} The above thermodynamic approach is quite different from that of Hill. \cite{Hill} The phase equilibrium is not given by Eq.~(6) as is assumed by Hill, but by Eq.~(4). This allows us to introduce, in the thermodynamic relationships, the elastic energy of the solid besides the surface free energy that is to connect the deformation of the solid to the variation of the surface free energy. In the approach of Hill, the surface term is not separated from the perturbation undergone by the bulk of the solid. Moreover, this perturbation is not clearly depicted since the variation of the volume of the solid induced by adsorption, i.e., the elastic energy, is not taken into account.
 \subsubsection*{\textit{II-Hysteresis region}}
 \label{subsubsecII:hyst reg}
\begin{figure}
\resizebox{\columnwidth}{!}{\includegraphics{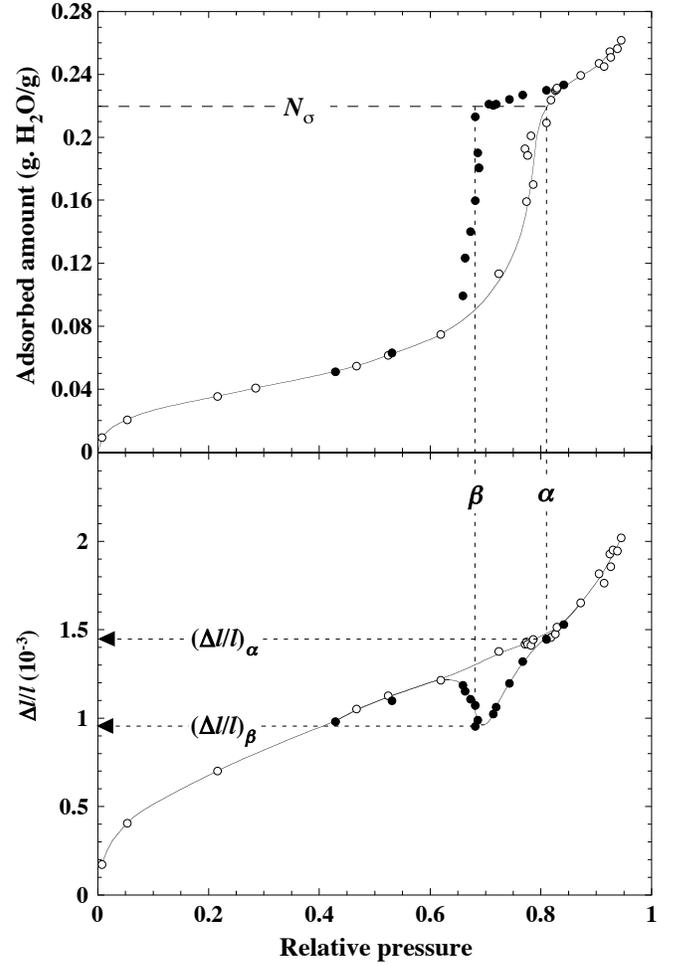}} 
\caption{Adsorption isotherm of water in porous glass at $18.75$~$^{o}$C and the corresponding linear extension measured during the adsorption-desorption cycle by Amberg and McIntosh (Ref.~\cite{Amberg52}). For a given adsorbed amount, equal to $0.22$~g.H$_{2}$O/g.porous glass, we have represented the two corresponding states of the system, $\alpha$ [$P_{\alpha}/P_{0}$=$0.81$, $(\Delta l/l)_{\alpha}$=$0.145\times10^{-2}$] and $\beta$ [$P_{\beta}/P_{0}$=$0.68$, $(\Delta l/l)_{\beta}$=$0.096\times10^{-2}$].}
\label{fig:Amberg2}
\end{figure}
As shown above, the variation of the surface free energy is proportional to the deformation. Quinn and McIntosh were the first to use this fact to determine the variation of the surface free energy within the hysteresis loop by measuring the dimensional change of the solid; in fact this cannot be done any other way.

In the hysteresis region, it is interesting to estimate the variation of $F_{\sigma,\rm so}$ between two states $(\alpha,\beta)$ corresponding to the same adsorbed amount, $N_{\sigma}$, on the two branches of the hysteresis loop as has been done by Quinn and McIntosh for the porous glass-water and porous glass-butane \cite{Quinn57} systems. The sign of this variation indicates which of these states is the more stable.

For such an estimate, we use the data of Amberg and McIntosh represented in Fig.~\ref{fig:Amberg2}, which correspond to water adsorption isotherm data in a porous glass rod together with the associated variation of the rod length. We consider two states $(\alpha,\beta)$ at the top of the hysteresis loop. Let us recall that in this region, the adsorption is reversible just as the marked observed contraction so that no capillaries empty. It has been suggested that this contraction is due to the increased value of negative pressure as a result of the changing radii of concave menisci at the outer region of the porous solid. \cite{Amberg52}

According to Eq.~(16) we have: 
\begin{equation}
\Big|F_{\sigma,\rm so}\Big|_{\alpha}^{\beta} =\Big|-P_{L}V_{L}\Big|^{\beta}_{\alpha}+k_{B}TN_{\sigma}\,ln\bigg(\frac{P_{\beta}}{P_{\alpha}}\bigg)-k \bigg|\frac{\Delta l}{l}\bigg|^{\beta}_{\alpha},
\end{equation}
where $P_{L}$ is the liquid pressure. 

The term $k(\Delta l/l)$ concerns the solid-liquid interface. Rigorously, we must take into account a new interface, the interface between the liquid which fills the pores and the vapor, by introducing the term $\gamma_{\rm lv}\big|A_{m}\big|_{\alpha}^{\beta}$, where $\gamma_{\rm lv}$ is the surface tension at the liquid-vapor interface and $A_{m}$ is the surface area of the menisci. The order of magnitude of the surface area of the menisci is given by the external surface area of the porous material. In Amberg and McIntosh experiment, the porous glass rod used was $0.73$~cm in diameter and $11.1$~cm in length, which corresponds to an external surface area of $2.1$~cm$^{2}$/g. As $\gamma_{\rm lv}\approx 70$~mJ/m$^{2}$, $\gamma_{\rm lv}\big|A_{m}\big|_{\alpha}^{\beta}$ is of the order of $10^{-5}$~J/g which is extremely small compared to the two last terms of Eq.~(34) as shown below.

The estimation of the first term is not so easy. The pores are full of liquid submitted to negative pressures due to the presence of concave menisci at the liquid/vapor interface. The problem is how to estimate $P_{L}$.

In a porous solid assumed to be inert, the liquid pressure is classically calculated as follows. Two equilibrium conditions are formulated simultaneously. One mechanical given by Laplace's law
\begin{equation}
P-P_{L}=2\,\frac{\gamma_{\rm lv}}{r_{m}}
\end{equation}
where $\gamma_{\rm lv}$ is the surface tension at the liquid/vapor interface and $r_{m}$ is the radius of the menisci, the other chemical given by Eq.~(6) if the porous solid is supposed to be inert. The Laplace-Kelvin equation resulting from these two equilibrium conditions gives $P_{L}$ as a function of $P$: 
\begin{equation}
P_{L}-P=\frac{k_{B}T}{v_{L}}\,ln\bigg(\frac{P}{P_{0}}\bigg)\approx P_{L},
\end{equation}
where $v_{L}$ is the molecular volume of the liquid.

As we have already noted in Sec. I, the hysteresis loop observed in porous glass is explained so far by the so-called pore blocking model. At the top of the hysteresis loop the porous system cannot empty as long as the constrictions located at the outer surface of the porous system are full of liquid. As the vapor pressure is reduced, menisci are formed in these constrictions, all the menisci having the same radius. Thus, if the emptying of porous glass is indeed controlled by pore blocking, the liquid pressure should obey Eq.~(36) as long as the porous system is full of liquid. According to Eq.~(36), $P_{L}$ should be, at a given vapor pressure, $5.9$ times higher for water than for butane.

Hooke's law tell us that the deformation is proportional to the pressure: 
\begin{equation}
\Delta P_{L} = -3K (\Delta l/l)
\end{equation}
where $K$ is the bulk modulus. Thus, in the same vapor pressure range, ($\Delta l/l$) should be also $5.9$ times higher for water than for butane. It is not the case. Indeed, if we consider the results shown in Fig.~\ref{fig:Amberg2} (water at $18.75^{o}$C) together with those of Ref.~22 (butane at $-6.2^{o}$C), we find that along the reversible region at the top of the hysteresis loop, the relationship ($\Delta l/l$) vs $ln(P/P_{0})$ is practically linear but the slope is only $1.9$ times higher for water than for butane.

This leads to an embarrassing result: the K value is found to depend markedly on the adsorbate: it is $3$ times higher for water than for butane. In those days, these results were discussed at length. According to Quinn and McIntosh, this discrepancy could be due to "a different stress distribution in the solid in the (two) cases owing possibly to different distributions of adsorbate." The question is why this occurs only when the pores are full of liquid. On the contrary, when the pores are full of liquid, there should be a nearly uniform stress distribution throughout the adsorbent. The idea of Quinn and McIntosh was also criticized by Sereda and Feldman, \cite{Sereda67} who proposed that "perhaps the real reason for the discrepancy is that the concept of changing curvature of the menisci is not valid." 

In fact, the above results can be explained if we take into account the elastic deformation of the porous matrix. $P_{L}$ is not given by Eq.~(36) because the chemical equilibrium of the system is governed by Eq.~(12) and not by Eq.~(6). Concerning the mechanical equilibria, in addition to the liquid-vapor interface [Eq.~(35)], we must take into account the solid-liquid interface. For a cylindrical interface of radius $R$, Laplace's law can be written as
\begin{equation}
\tau_{n}-P_{L}=\frac{\gamma_{\rm sl}}{R},
\end{equation}
where $\tau_{n}$ is the bulk stress along the normal to the interface and $\gamma_{\rm sl}$ is the surface tension at the solid-liquid interface. Thus, through Eq.~(38), $P_{L}$ depends on the state of deformation of the solid and through Eq.~(35), $r_{m}$ depends on $P_{L}$. The radius of the menisci can be controlled through the deformation of the porous solid to homogenize, for example, the liquid pressure in the porous material if the pores are noninterconnected as it is the case, e.g., in porous silicon.

A last remark concerning the experiment of Quinn and McIntosh. The linear deformation is significantly higher for water adsorption than for butane adsorption which suggests that the deformation increases with the surface tension ($\gamma_{\rm lv}\simeq70$~mJ/m$^{2}$ for water and $\simeq15$~mJ/m$^{2}$ for butane \cite{Miqueu}). This was also observed by Herman et al. \cite{Herman06} in silica aerogel.

We consider the results of Quinn and McIntosh as the first proof that the evaporation of fluid from porous glass is not governed by pore blocking.

Finally, we cannot use Eq.~(36) to determine the term $\big|-P_{L}V_{L}\big|^{\beta}_{\alpha}$ in Eq.~(34), but we know that $P_{L}$ decreases with the vapor pressure $P$ since the relative extension ($\Delta l/l$) decreases with $P$ so that this term is positive.

The two last terms of the right member of Eq.~(34) can be easily calculated. We obtain
\begin{equation}
\Big|F_{\sigma,\rm so}\Big|_{\alpha}^{\beta}\rm(J/g.adsorbent)=\Big|-P_{L}V_{L}\Big|^{\beta}_{\alpha} -5.14\,+7.06.
\end{equation} 

Thus, the free energy difference at the top of the plateau is positive and at least equal to $(7.06-5.14)$~J/g.adsorbent.

Consider now the case of porous Si. We have no information on the deformation of porous silicon for N$_{2}$ adsorption but we know that in the hysteresis region, the deformations of porous material depend on the negative pressure of the liquid and thus, through Eq.~(35), on the surface tension $\gamma_{\rm lv}$. For pentane at ambient temperature, \cite{Dolino96} $\gamma_{\rm lv}\simeq14$~mJ/m$^{2}$, while for nitrogen, \cite{Gregg82} $\gamma_{\rm lv}\simeq9$~mJ/m$^{2}$. Thus, we can expect that for nitrogen adsorption, the deformation is of the same order of magnitude as for pentane. We can repeat the calculations represented by Eq.~(34) for the isotherm of Fig.~\ref{fig:hystbehavior} where $(P/P_{0})_{\beta}=0.65$ and $(P/P_{0})_{\alpha}=0.8$. We find $k_{B}TN_{\sigma}\,ln\bigg(\frac{P_{\beta}}{P_{\alpha}}\bigg)=-1.64$~J/g (the mass of the porous solid equals $36.36\times10^{-3}$~g).

From Eq.~(31), we can estimate the proportionality factor $k$. As we have seen above, in the case of porous silicon layers, the contraction or the extension is unilateral. Thus $\alpha=1$ and, according to classical theory of elasticity, \cite{Landau} the constant $C$ defined by Eq.~(22) is related to Young's modulus $E_{p}$ by the relation
\begin{equation}
C=E_{p}\frac{1-\nu_{p}}{1-\nu_{p}-2\nu_{p}^{2}},
\end{equation} 
where $\nu_{p}$ is the Poisson's coefficient of the porous layer. Young's modulus $E_{p}$ for porous silicon can be estimated by the Gibson and Ashby \cite{Gibson} relation $E_{p}=E~(1-P_{or})^{2}$, where $E=166$~GPa is Young's modulus for silicon and $P_{or}$ is the porosity of the layer. In our case, $P_{or}\simeq0.5$ so that $E_{p}\simeq40$~GPa. For a porous sample \cite{Barla} similar to that under study, $\nu_{p}=0.09$ so that $C\simeq E_{p}$. For pentane, the relative contraction of the lattice parameter of the porous layer perpendicular to the substrate, $(\Delta a/a)_{\bot}$, measured in the hysteresis region, equals a few $10^{-4}$. Taking $(\Delta a/a)_{\bot}=10^{-4}$ and a density of $2.32$~g/cm$^{3}$ for silicon, we find $k\Big|(\Delta a/a)_{\bot}\Big|_{\alpha}^{\beta}=1.72$~J/g. We see that, as for porous glass, the variation of the surface free energy caused by the deformation of the porous material is an important component of the total free energy.

We note, concerning the assumption of inert adsorbents, that if the solid is supposed to be inert, we have
\begin{equation}
\Big|F_{\sigma}\Big|_{\alpha}^{\beta} =\Big|-P_{L}V_{L}\Big|^{\beta}_{\alpha}+k_{B}TN_{\sigma}\,ln\bigg(\frac{P_{\beta}}{P_{\alpha}}\bigg)+\gamma_{\rm lv} \big|A_{m}\big|_{\alpha}^{\beta}\,
\end{equation}
where $P_{L}$ is given by Eq.~(36). In this case, $\big|-P_{L}V_{L}\big|^{\beta}_{\alpha}=-k_{B}T\,N_{\sigma}\, ln\big(P_{\beta}/P_{\alpha}\big)$ so that $\big|F_{\sigma}\big|_{\alpha}^{\beta}\simeq \gamma_{\rm lv} \big|A_{m}\big|_{\alpha}^{\beta}$. We found that $\gamma_{\rm lv} \big|A_{m}\big|_{\alpha}^{\beta}\simeq10^{-5}\,-\,10^{-4}$~J/g for porous glass and porous silicon. Thus, the difference of the free energy between the two branches of the hysteresis loops would be extremely small compared to the two first terms of Eq.~(41). The fluctuations of each of these two terms in adsorption experiments are certainly higher that $10^{-5}\,-\,10^{-4}$~J/g so that the hysteresis phenomenon should be unobserved for these porous materials. 

 At the end of this section we summarize a few important conclusions which can be extended to all the porous materials:

i) The surface free energy is directly related to the elastic deformation of the solid and is a major component of $F_{\sigma, \rm so}$, the total free energy of the two-component system.

ii) The variation of $F_{\sigma,\rm so}$ between two states $(\alpha,\beta)$ corresponding to the same adsorbed amount, $N_{\sigma}$, on the two branches of the hysteresis loop is positive. This shows that the condensation branch represents the more stable states and that a barrier exists to the emptying of the porous system, the height of the barrier depending essentially on the solid deformation.

In the following section, we describe qualitatively how the pores can interact during the adsorption-desorption process through the pore wall deformation.
\subsection{A qualitative description of the interaction between the pores during the adsorption-desorption process through the pore wall deformation}
\label{subsec:qualitative}
We have seen in Sec.~\ref{subsec:elasdef} that the deformation of porous materials during the condensation-evaporation process presents common features. The amplitude of these deformations depends evidently on the stiffness of the porous material. The stiffer the material, the lower the amplitude of the deformation. Intuitively, we could believe that the effect of the elastic deformation on the adsorption-desorption process would be small in stiff material and large in weak material. It is not the case. Actually, in Sec.~\ref{subsec:thermo}, we have shown that the crucial parameter is the surface free energy, which is proportional to the linear elastic deformation and to the elastic constant of the solid, so that the effect of which we speak--the surface free energy change--can be as important in stiff material such as porous silicon or porous glass as in weak material such as aerogel. 

At this stage, we must specify the limits of the following discussion. It concerns porous materials for which:

(1) the notion of pores separated by walls makes sense, which is not the case of the low density aerogels \cite{Detcheverry, Bonnet} which rather consist of void in which some small solid particle impurities have aggregated. 

(2) the change in pore size or volume upon condensation or evaporation can be neglected which is not the case of most of aerogels. \cite{Scherer01} This does not mean that the effect that we analyse in this paper does not exist in these materials--the thermodynamics approach developed in Sec.~\ref{subsec:thermo} is generic--this means that this effect is in addition to others which complicates the analysis. For example, the analysis of subloops inside the boundary hysteresis loop which is essential to determine whether a porous material is composed of independent pores or not is rendered problematic if the change in pore size cannot be neglected.

In the present section, we describe qualitatively how fluid molecules adsorbed in neighboring pores can interact through the elastic deformation of the pore walls. Basing on this interaction mechanism and on the hysteretic behavior shown in Fig.~\ref{fig:hystbehavior}, we then propose a qualitative description of the pore filling and emptying.
\subsubsection{Interaction mechanism through pore wall elastic deformation}
The linear relationship between the variation of the surface free energy and the elastic deformation of the solid [Eq.~(32)] is the result of equilibrium conditions. The first, given by Eq.~(12) corresponds to chemical equilibrium, the second, given by Eq.~(24), corresponds to mechanical equilibrium. Adsorption or capillary condensation of gas in a given pore causes elastic deformation not only of its inner pore walls but also of the inner walls of its neighbors and thus causes a change of their surface free energy.
\begin{figure}
\resizebox{\columnwidth}{!}{\includegraphics{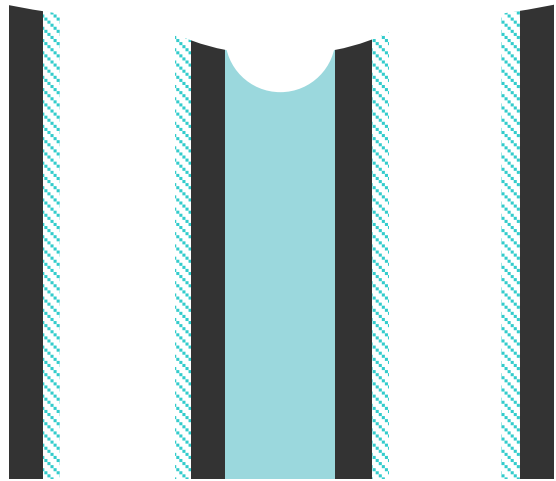}} 
\caption{(Color online). Schematic representation of the deformation of the pore walls (in black) surrounding a filled pore with a concave liquid meniscus. The negative pressure of the liquid in the filled pore contracts its inner walls and reduces the extension of the inner walls of the neighboring empty pores submitted to the spreading pressure exerted by an adsorbed film (hatched zones).}
\label{fig:Schempore}
\end{figure}
Figure~\ref{fig:Schempore} gives, for example, a picture of the interaction between a filled pore with a meniscus and surrounding larger empty pores with an adsorbed film on their walls. The negative pressure of the liquid in the filled pore tends to contract its inner walls whereas the inner walls of the neighboring empty pores are submitted to the spreading pressure exerted by the adsorbate. The forces exerted on both sides of the pore walls create elastic stresses in the solid, which reduce both the contraction of the inner walls of the filled pore and the extension of the inner walls of the empty pores. Thus, the surface free energy of a pore, and hence the adsorbed amount if it is empty (see Figs.~\ref{fig:AmbergGamma} and \ref{fig:AmbergAds}) or the liquid pressure, according to Eqs.~(35) and (38), if it is filled, depends on the state of the neighboring pores.
\subsubsection{Scenario for the pore filling and emptying}
\begin{figure}
\resizebox{\columnwidth}{!}{\includegraphics{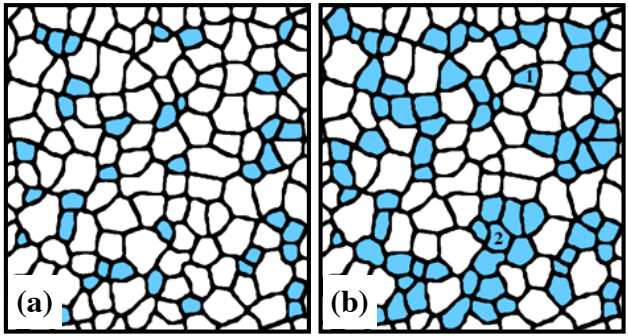}} 
\caption{(Color online). Binary image of a bright field transmission electron microscopy plane view of a porous Si layer obtained by reproducing both the porosity of the layer and the consistency in the wall thickness (see Ref.~2 for more details). The empty pores (in white) and the filled pores (in gray) are separated by Si walls (in black). Panels (a) and (b) represent different steps for the adsorption process, the vapor pressure increasing from panel (a) to (b). In panel (b) we have pointed out two filled pores of similar size, number 1 with neighboring empty pores and number 2 with neighboring filled pores.}
\label{fig:stepAds}
\end{figure}

\begin{figure}
\resizebox{\columnwidth}{!}{\includegraphics{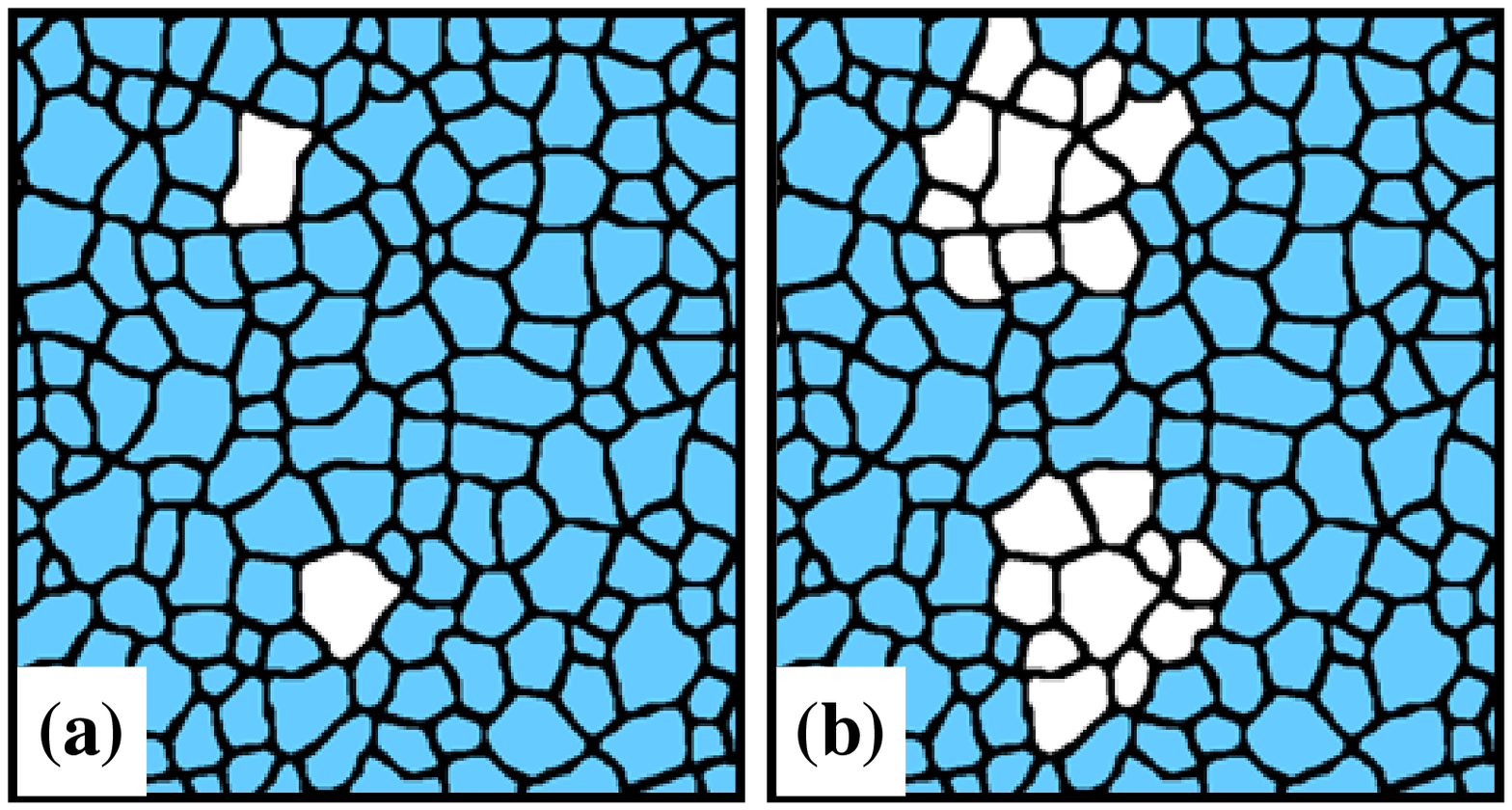}} 
\caption{(Color online). Schematic representation of the beginning of the evaporation viewed from the same binary image such as that shown in Fig.~\ref{fig:stepAds}. The vapor pressure decreases from panel (a) to (b). Panel (a) shows two pores (in white) in which the evaporation is initiated. Panel (b) shows the evaporation of the nearest neighboring pores in a single cooperative event.}
\label{fig:stepDes}
\end{figure}

Based on this interaction mechanism, we give here a qualitative description of the pore filling and emptying. We take p$^{+}$-type porous silicon as an example, the morphology of which is pretty well known and has the advantage of being easily and accurately schematized as a 2D image \cite{CoasnePRL,Grosman08} such as those shown in Figs.~\ref{fig:stepAds} and \ref{fig:stepDes}.

The beginning of the condensation process is illustrated in Fig.~\ref{fig:stepAds}(a) where the smallest pores are filled. They are randomly distributed in the porous matrix, most of them being surrounded by larger pores still empty. The walls of the neighboring empty pores contract compared to their states before the filling of the central pore, their surface free energy increases, resulting in a decrease of the adsorbed amount (see Fig.~\ref{fig:AmbergAds}). Thus, the presence of a filled pore tends to delay the filling of its neighbors towards higher pressure or at least does not favor their filling. During the condensation process the filled pores are not grouped but rather randomly scattered. The order according to which the pores fill is not perturbed by the interaction mechanism: they fill sequentially from the narrowest to the largest. This is supported by the fact that the extension of the condensation branch on the pressure axis is representative of the pore size distribution. As shown below, it is not the case for evaporation.

\begin{figure}
\resizebox{\columnwidth}{!}{\includegraphics{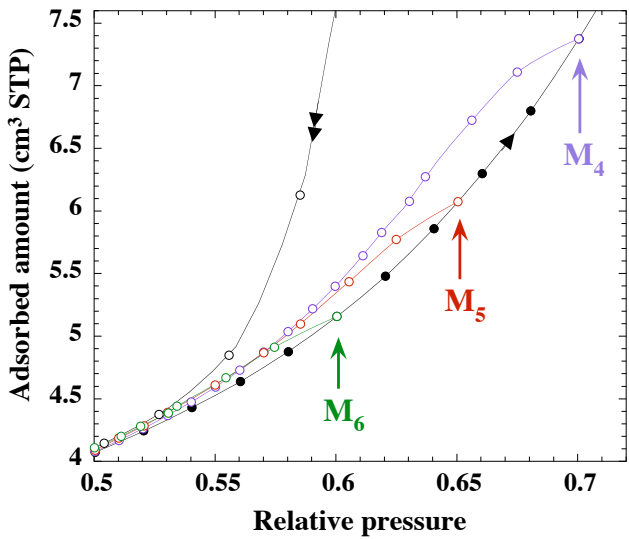}} 
\caption{(Color online). Magnification of the hysteresis loop region of the nitrogen adsorption isotherm shown in Fig.~\ref{fig:hystbehavior}, together with three PDSCs starting from points $M_{4}$, $M_{5}$ and $M_{6}$ on the boundary condensation branch. As the reversal point approaches the lowest closure point of the hysteresis loop, the hysteresis loop shrinks.}
\label{fig:PDSC}
\end{figure}

In the beginning of the condensation process, the average distance between the filled pores is the longer as they are fewer. The interaction between them can be neglected. They do not "see" each other. The PDSCs shown in Figs.~\ref{fig:hystbehavior} and \ref{fig:PDSC} indicate that as the reversal point M on the condensation branch approaches the lowest closure point of the hysteresis loop, the hysteresis phenomenon is less and less important. These two observations suggest that the emptying of filled pores surrounded by empty pores and located far from other filled pores is quasireversible.
 
As the vapor pressure is increased, larger pores, in various environments, fill. Figure~\ref{fig:stepAds}(b) illustrates two extreme cases: two filled pores of similar size, number 1 with neighboring empty pores, number 2 with neighboring filled pores. The walls of the filled pore are less contracted in the first case than in the second. Clearly, the surface free energy of these two pores is different and this can change their evaporation pressure. The PDSCs shown in Fig.~\ref{fig:hystbehavior} indicate that the hysteresis phenomenon becomes increasingly important as the reversal point tends to the highest closure point of the hysteresis loop. Since, as we saw previously, a filled pore surrounded by empty pores fills and empties quasireversibly, this suggests that the emptying of a pore surrounded by filled pores is blocked by the presence of the neighboring filled pores. This idea is supported by the calculations of the previous section which show that when the porous material is completely filled, there is an energy barrier to evaporation, the height of which depends essentially on the elastic deformation.

Consider now the boundary evaporation branch. As the pressure is decreased from C to D (Fig.~\ref{fig:schemDeltaL}), the porous material contracts under the effect of the negative liquid pressure created under the concave menisci at the top of the pores. As one further decreases the pressure, the porous system begins to empty.

At this stage, very little is known about the exact manner in which the pores empty. According to the pore blocking model and its derivatives, a concave meniscus is formed at the liquid-vapor interface and recedes from the interface to the inside of the porous solid, at thermodynamic equilibrium. In p$^{+}$-type porous silicon, this model has been ruled out. \cite{CoasnePRL,Wallacher04,Grosman08,GrosmanCav} For example, the experiment of Wallacher \textit{et al.} \cite{Wallacher04} shows that a porous layer can drain across layers, several microns thick, full of liquid. In porous glass, the experimental observations of Quinn and McIntosh do not match this model either. This suggests that the evaporation of fluid from porous materials is either via the formation of gas bubbles or via the sudden propagation of the menisci leading to evaporation out of equilibrium. This will be discussed in detail in a future paper. \cite{GrosmanCav}

What is known is that the evaporation branch is steeper than the condensation branch, which indicates that the evaporation occurs in an avalanchelike manner. Some pores begin to empty [Fig.~\ref{fig:stepDes}(a)], leading to the extension of the neighboring pore walls. The neighboring filled pores are now in contact with empty pores which, as we have seen above, facilitates their draining, leading to the emptying of some of them in a single cooperative event [see Fig.~\ref{fig:stepDes}(b)]. Note that, in Fig.~\ref{fig:stepDes}(b), the distribution of the empty pores, around the two pores at the origin of the evaporation, is schematic, the exact manner in which the empty pores "propagate" being probably more complicated. 

Avalanche phenomena have been experimentally observed in draining of superfluid helium from Nuclepore. \cite{Lilly1, Lilly2} The authors attributed this behavior to the presence of both the mobile helium film and pore-pore intersections. However, since the presence of pore-pore intersections has no significant role in the evaporation process, their presence is probably not relevant in these avalanche phenomena. To confirm this idea it would be interesting to carry out such experiments in porous silicon.

Pore-space correlation in porous glass has been studied using ultrasonic attenuation and light scattering. \cite{Page93, Page95} These studies show that the distributions of the empty pores are quite different on filling and on draining. During the condensation process, no long-range correlation of the empty pores are observed, while, on drainage, the empty pores exhibit long range correlations with a fractal structure which, according to the authors, suggests that the drainage can be modeled by analogy to invasion percolation. We have seen above that in porous glass, the experimental observations of Quinn and McIntosh do not match a pore blocking/percolation process. According to our wall elastic deformation model, the emptying of pores occurs from neighbor to neighbor, a mechanism which seems difficult to distinguish, at a first glance, from percolation. It is the reason why the hysteretic behavior simulated by Mason \cite{Mason} resembles qualitatively the experimental hysteretic behavior of porous silicon.
\section{Conclusion}
It has been known for a long time that the adsorption and condensation of gas induce elastic deformation of the porous matrix. The reversible formation of an adsorbed film, which preceeds capillary condensation, results in an extension of the porous material while, when the pores are full of liquid, a marked contraction occurs due to large negative liquid pressure generated by the formation of concave menisci. The elastic deformation exhibits an hysteresis loop in the same pressure region as the adsorption phenomenon.

In the present paper, we develop a thermodynamic approach which takes into account these elastic deformations. In addition to the adsorbate and the adsorbate-solid interface, the two-component system we consider includes the elastic energy stored in the solid during adsorption, a parameter which has been always ignored in the physics of adsorption.

Our theoretical approach is developed in parallel to the experimental work of Amberg and McIntosh, \cite{Amberg52} who measured both the length change of a porous glass rod and the adsorbed amount during adsorption of water. In the region of reversible adsorption, before capillary condensation occurs, they found that the relative extension of the rod is proportional to the decrease of the surface free energy. 

Taking into account the elastic energy of the solid, we explain why the variation of the surface free energy depends on the deformation and on the elastic constants of the porous solid. The surface free energy is an important component of the total free energy: the assumption of an inert solid in physics of adsorption is totally unjustified.

In the hysteresis region, we show that the condensation branch represents the more stable states and that an energy barrier exists to the evaporation, the height of which depends essentially on the elastic deformation of the substrate.

Since the surface free energy of pores is directly related to the deformation of their inner walls, the fluid inside them interacts through the deformation of the pore walls. Based on this interaction mechanism and on the shape of the descending scanning curves which are similar for all the studied porous systems, we propose a scenario for the filling and emptying of pores.

During the condensation process, the interaction mechanism does not modify the order according to which the pores fill, i.e., from the smallest to the largest. The emptying of a filled pore surrounded by empty pores is quasireversible and becomes irreversible when it is surrounded by filled pores. This provides an explanation for the steepness of the boundary evaporation branch. When the porous system begins to empty, the filled pores in contact with empty pores will first empty, leading to the emptying of their neighbors and then of the whole system in an avalanchelike manner. The irreversibility of the adsorption-desorption process in porous media would not originate in the irreversibility of the process in individual pores but would be a property of the whole system.

MCM-41 and SBA-15 porous materials exhibit boundary hysteresis loops of type H1, with steep and parallel condensation and evaporation branches while porous silicon and porous glass present hysteresis loop of type H2 with a broad condensation branch and a steep evaporation branch. Two different mechanisms for evaporation were attributed to these two hysteresis loops: evaporation at thermodynamic equilibrium for H1 type, pore blocking/percolation process for H2 type. The present paper and previous ones show the fallacy of this classification: the cylindrical pores of SBA-15 and MCM-41 do not empty at equilibrium, \cite{Morishige04,Grosman05} and there is no pore blocking/percolation process in porous silicon \cite{Grosman08} and in porous glass. Hysteresis loops of type H1 and H2 could be simply characteristic of a narrow pore size distribution (PSD) and of a large PSD, respectively, whatever the shape of the pores is and whether they are connected or not.

The interaction mechanism we propose acts as an alternative to the percolation process. The coupling parameter is the surface free energy which is transmitted from a pore to its neighbors through the elastic deformation of the pore walls. This coupling parameter, which is common to all porous materials, allows us to explain why the pores interact whether they are interconnected or not.
\begin{acknowledgments}
We acknowledge the Canadian Journal of Chemistry for granting us the permission to use the work of C. H. Amberg and R. McIntosh. The authors are grateful to Geoffroy Pr\'evot for fruitful discussions.
\end{acknowledgments}

\end{document}